\input{aipcheck}

\documentclass[
  ,draft            % use draft while you are working on the paper
  ]
  {aipproc}

\layoutstyle{6x9}

\begin{document}

\title{New Tools for Probing the Phase Space Structure of Dark Matter Halos}


\author{Monica Valluri}{
  address={University of Michigan}
}

\author{Victor P. Debattista}{
  address={University of Central Lancashire}
}

\author{Thomas Quinn}{
  address={University of Washington}
}

\author{Ben Moore}{
  address={University of Z\"urich}
%  ,altaddress={<author1 address>} % additional visiting address
}

\begin{abstract}
We summarize recent developments in the use of spectral methods for analyzing large numbers of orbits in $N$-body simulations  to obtain insights into the global phase space structure  of  dark matter halos. The fundamental frequencies of oscillation of orbits can be used to understand the physical mechanism by which the shapes of dark matter halos evolve in response to the growth of central baryonic components. Halos change shape primarily because individual orbits change their shapes adiabatically in response to the growth of a baryonic component, with those at small radii become preferentially rounder. Chaotic scattering of orbits occurs only when the central point mass is very compact and is equally effective for centrophobic long-axis tube orbits as it is for centrophilic box orbits. 
\end{abstract}

\maketitle

%%%%%%%%%%%%%%%%%%%%%%%%%%%%%%%%%%%%%%%%%%%%
%% MAINMATTER
%%%%%%%%%%%%%%%%%%%%%%%%%%%%%%%%%%%%%%%%%%%%

\section{Introduction}

While collisionless $N$-body simulations produce dark matter halos that are triaxial or prolate, simulations which include the effects of baryons result in more spherical or axisymmetric halos\citep{dubins_94, kkzanm04,deb_etal_08}. It has been suggested that the change in shape could be the consequence of chaotic scattering of the box orbits that form the "back bone" of triaxial galaxies \citep{ger_bin_85}. Alternatively,  halo shapes might change due the adiabatic response of orbits to the change in the central potential. In a recent paper \citet{valluri_etal_10}  showed that by applying a spectral method to analyze large numbers of randomly selected orbits in $N$-body simulations it was possible to clearly distinguish between these two options. In this paper we present a brief summary of some of their main results.
 
\section{Frequency Analysis of $N$-body orbits} 

Prolate and triaxial dark matter halos were formed from the merger of spherical NFW halos \citep{nfw} and  baryonic components (representing a disk, an elliptical galaxy or a massive compact nucleus) were grown adiabatically with time. The evolution of the halo was followed using {\sc pkdgrav} an  efficient, multi-stepping, parallel tree code \citep{stadel_phd}.  After the baryonic component was grown to full strength it was  artifically "evaporated" to  allows us to test for the importance of chaos in the evolution of the system \citep[][hereafter D08 and references therein]{deb_etal_08}.
In each system several thousand orbits were selected and their trajectories evolved in a frozen potential in each phase of the evolution of each halo. The initial triaxial/prolate phase is referred to as {\it phase a}, once the baryonic component is grown to full strength the halo is in {\it phase b}, and after the baryonic component had "evaporated" and the system returned to equilibrium the halo is in {\it phase c}. Each orbit was analyzed using a code that decompose its phase space trajectory to obtain its three fundamental frequencies of oscillation \citep{laskar_90, valluri_merritt_98}.  
The fundamental frequencies were then used to obtain a complete picture of the properties of individual orbits: to distinguish between regular and chaotic orbits,  to classify regular orbits into major orbit families, to quantify the average shape of  an orbit and relate its shape to the shape of the halo, and to identify the major resonant families of the halo which determine its structure \citep{valluri_etal_10}. 

\section{Results}
%\subsection{Extended vs. centrally concentrated baryonic components}

\begin{figure}
\centering
\includegraphics[width=.37\textwidth]{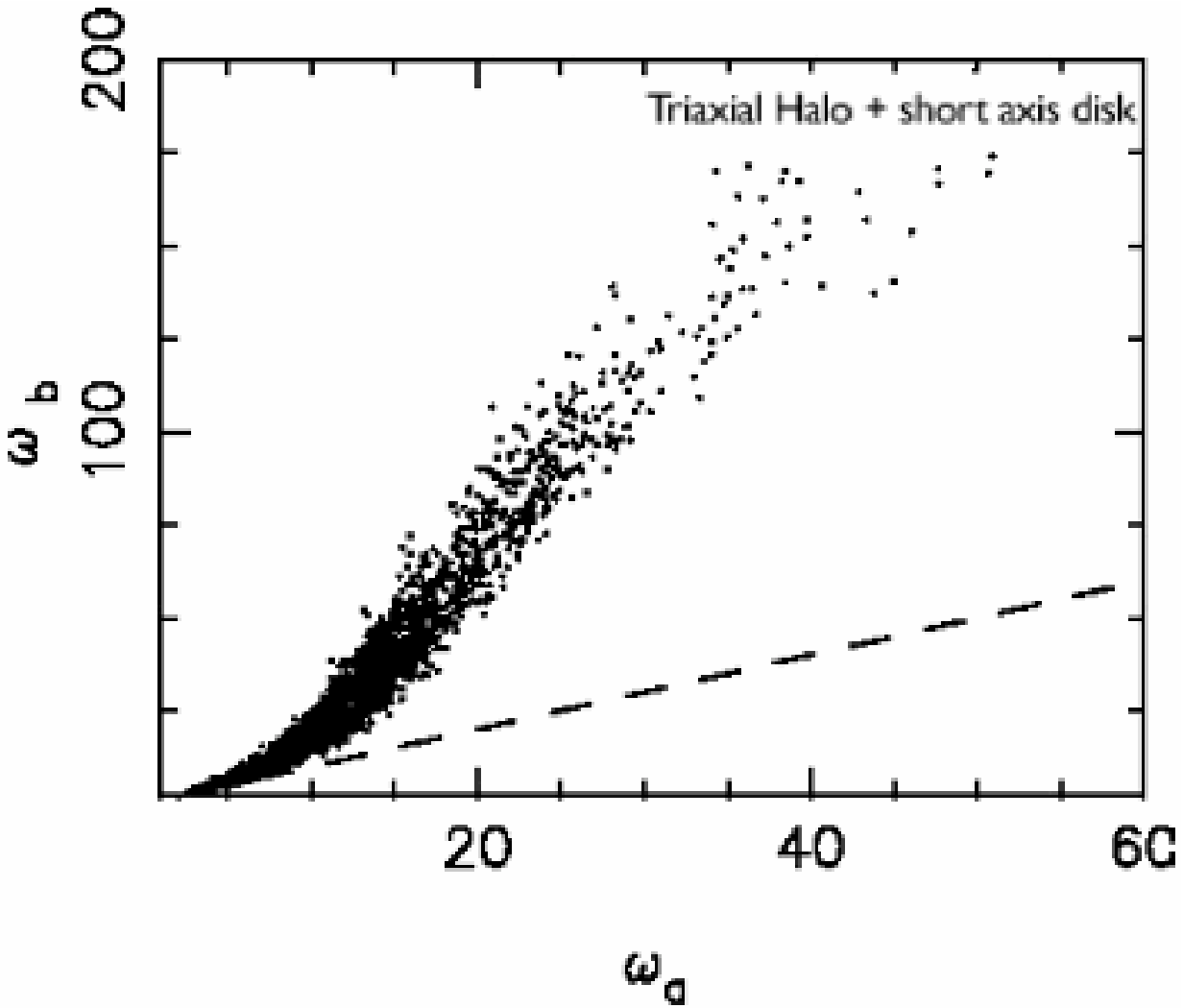}
  \hfill
\includegraphics[width=.37\textwidth]{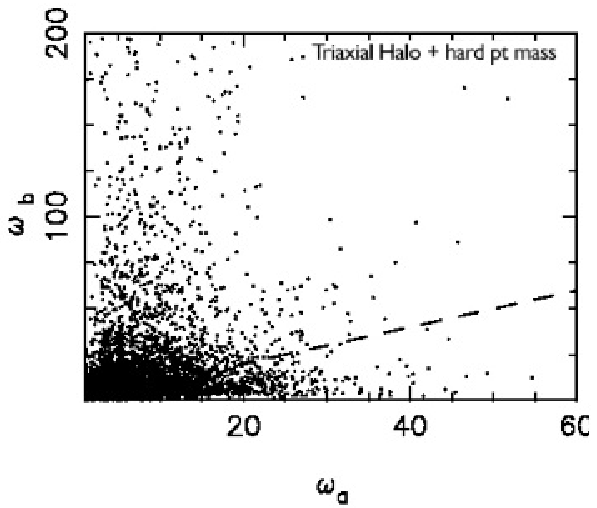}
\caption{For $\sim$5000 particles frequencies before ($\omega_a$)  versus after ($\omega_b$) the growth of a baryonic component are shown. {\it Left:} effect of an extended baryonic component (disk with 3~kpc scale length). {\it Right:} effect of a hard spherical point masses with 0.1~kpc softening.
\label{fig:omega_abc}}
\end{figure}

The largest of the three fundamental frequencies in each of the three phases of the evolution of our models (referred to as $\omega_a$,  $\omega_b$ and $\omega_c$ respectively), can be compared to distinguish between adiabatic and chaotic evolution. In the case of a primarily adiabatic response, particles deep in the potential (large $\omega_a$) are expected to experience a greater increase in frequency than particles further from the center.  In Figure~\ref{fig:omega_abc} we compare the frequencies of orbits in an initially triaxial halo ($\omega_a$) with their frequency in the presence of a baryonic component ($\omega_b$).  For the extended disk ({\it Left}) (as well as other extended baryonic components), $\omega_b$ increases monotonically with $\omega_a$ with fairly small scatter, indicating that the orbits in this potential responded adiabatically (the dashed line shows the 1:1 correlation between the two frequencies). 
%%%% FIGURE 1 %%%%%
%%%% FIGURE 1 %%%%%
However, a hard central point mass ({\it Right}) results in  significant scattering in $\omega_b$ with small values of $\omega_a$ (i.e. weakly bound orbits) having some of the largest values
of $\omega_b$.  $\omega_b$ sometimes decreased instead of increasing providing further evidence for chaotic scattering in this case. 

The orbital frequencies were used to distinguish between regular and chaotic orbits and to  classify the regular orbits into major families. We studied both triaxial and prolate halos and our orbital analysis  showed that  the initial triaxial halo was composed of 84-86\%  box orbits, 11-12\% long-axis tube orbits, 2\%  short axis tubes, and  1-2\% chaotic orbits. In contrast the prolate halo had 15\% box orbits, 78\% long-axis tubes, 7\% short-axis tubes and no chaotic orbits. To test the hypothesis that it is the box orbits that are most significantly scattered by a central point mass we define the fractional change $\Delta\omega_{ac}$ in the frequency of an orbit  from its value in the initial halo ($\omega_a$) to its value after the baryonic component was evaporated ($\omega_c$)  to measure the amount of chaotic scattering.

\begin{figure*}
\centering
\includegraphics[width=0.4\textwidth]{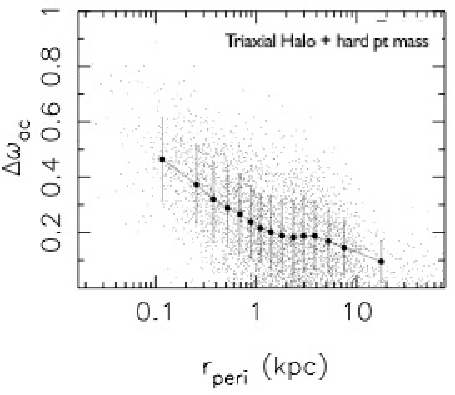}
\hfill
\includegraphics[width=0.408\textwidth]{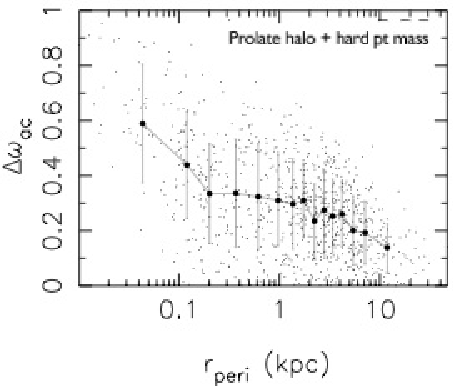}
%{figPlB3.pdf}
\caption{The change in frequency of an orbit $\omega_{ac}$ versus pericentric radii $r_{\rm peri}$ for  triaxial halo ({\it Left}) and prolate  halo ({\it Right}).
\label{fig:Pl3_jtot_rperi}}
% plots for this figure created by /kedara/shape/fmapnb/ver3/omegawa_wx_rp_j.f
\end{figure*}

 Figure~\ref{fig:Pl3_jtot_rperi}  shows that orbits with smaller pericenter radii $r_{\rm peri}$ experience a significantly larger change in frequency $\Delta \omega_{ac}$ than orbits at large pericenter radii. This is true for both the triaxial halo ({\it Left })  which is dominated by centrophilic box orbits and the prolate halo ({\it Right}) which is dominated by long-axis tubes. %We found that the change in frequency $\Delta \omega_{ac}$ was not correlated with any other quantity and in particular it was independent of orbital angular momentum.  There was no correlation with $r_{\rm peri}$ or any other quantity in models with extended bayonic components.  
 Thus chaotic scattering is equally strong for the centrophobic long-axis tube orbits and  centrophilic box orbits contrary to the prevailing view.

%%%% FIGURE 2 %%%%%
%%%% FIGURE 2 %%%%%

%\subsection{Evidence for orbital shape change}

\begin{figure*}
\centering
\includegraphics[width=0.33\textwidth,angle=-90]{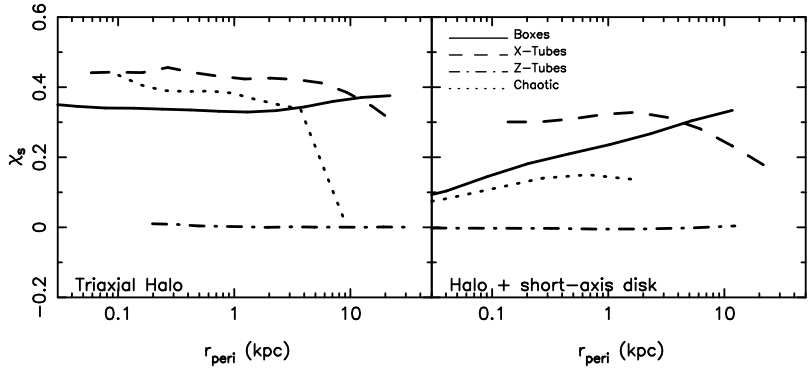}
%{orb5SA1_sh_rp.jpg}
\caption{The mean orbital shape parameter $\chi_s$ for different orbital types  (indicated by line legends) vs. pericentric radius $r_{\rm peri}$.}
\label{fig:shape_peri}
\end{figure*}

In a triaxial potential in which the major, intermediate and short axes are along the Cartesian directions   $x, y, z$ respectively, the oscillation frequencies satisfy the condition $
|\omega_x| < |\omega_y| < |\omega_z|$ for any orbit with the same over-all shape as the density distribution.  For such an orbit we can define a shape parameter $\chi_s \equiv  {|{\omega_y/\omega_z}|}  -  {|{\omega_x/\omega_z}|}$ %We can therefore define an orbit shape parameter $\chi_s$ by,
%
% \begin{eqnarray}
%   \chi_s       &\equiv {\frac{|\omega_y|}{|\omega_z|}} - {\frac{|\omega_x|}{|\omega_z|}} >  0.
% \end{eqnarray}
% 
which is positive for orbits with
elongation along the major axis of the figure, with larger values of $\chi_s$ implying a
greater degree of elongation.   
Since  orbits closer to the central potential are more significantly affected by
the baryonic component, we expect them to
become rounder ($\chi_s \rightarrow 0$) than orbits further out. Figure~\ref{fig:shape_peri} shows how the orbital shape parameter $\chi_s$  for various orbital types varies as a function of the pericenter radius $r_{\rm peri}$ for a triaxial halo ({\it Left}) and after a disk was grown in this halo with symmetry axis parallel to the short axis ({\it Right}).   In each plot the curves  show the mean shape distribution of orbits of a given orbital type at each radius, with orbital types indicated in the line-legends. In the initial triaxial halo boxes, long-axis tubes and chaotic orbits are significantly elongated  ($\chi_s \sim 0.35$) both at small and large radii. %(short axis tubes are always axisymmetric and show elongation of $\chi_s \sim 0$ along the major axis). 
After the growth of the short-axis disk the orbits at small radii become axisymmetric ($\chi_s \sim 0$) while orbits at large radii (especially boxes) remain quite elongated. This change in orbital shape at small radii was seen in all halos regardless of the radial scale length of the baryonic component.

%%%% FIGURE 3 %%%%%
%%%% FIGURE 3 %%%%%

\section{Implications}
The analysis of fundamental frequencies of orbits in $N$-body halos is a powerful technique that allows us to identify the primary physical processes that cause halo shapes to change in response to the growth of a baryonic component. We confirmed the conclusion reached by D08 that chaos is not an
important driver of shape evolution but found that significant
chaotic scattering does occur when the baryonic component is in the
form of a hard central point mass (of scale length $\sim 0.1$~kpc).
Regardless of the original orbital composition
of the triaxial or prolate halo, and regardless of the shape and radial
scale length of the baryonic component, halos become more oblate because orbits closer to
 the center of the potential become axisymmetric. Although the resultant halos are almost oblate, their orbit populations contain orbits  (boxes and long-axis tubes) that are characteristic of a triaxial rather than axisymmetric potential.

\begin{theacknowledgments}

 MV would like to thank the organizers of the conference at UCLAN and the University of Malta for organizing an excellent  meeting.
\end{theacknowledgments}



\bibliographystyle{aipproc}   % if natbib is available

\bibliography{master}

\IfFileExists{\jobname.bbl}{}
 {\typeout{}
  \typeout{******************************************}
  \typeout{** Please run "bibtex \jobname" to optain}
  \typeout{** the bibliography and then re-run LaTeX}
  \typeout{** twice to fix the references!}
  \typeout{******************************************}
  \typeout{}
 }

\end{document}